\documentclass{tADR2e}

\begin{document}

\title{Sensor-less Angle and Stiffness Control of Antagonistic PAM Actuator \\Using Reference Set}

\author{Takaya Shin$^{a}$$^{\ast}$\thanks{$^\ast$Corresponding author. Email: shintakaya@uec.ac.jp
\vspace{6pt}} and Kiminao Kogiso$^{a}$\\\vspace{6pt}  $^{a}${Department of Mechanical and Intelligent Systems Engineering, \\
The University of Electro-Communications, Tokyo, Japan}}

\maketitle

\begin{abstract}
This paper proposes a simultaneous control method for the angle and stiffness of the joint in an antagonistic pneumatic artificial muscle (PAM) actuator system using only pressure measurements, and clarifies the allowable references for the PAM actuator system.
To achieve a sensor-less control, the proposed method estimates the joint angle and contraction forces using an unscented Kalman filter that employs a detailed model of the actuator system. 
Unlike previous control methods, the proposed method does not require any encoder and force sensor to achieve angle and stiffness control of the PAM actuator system.
Experimental validations using three control scenarios confirm that the proposed method can control the joint angle and stiffness simultaneously and independently.
Moreover, it is shown that a reference admissible set can be used as an indicator to establish reference values by demonstrating that the reference set covers the experimentally obtained trajectories of the angle and stiffness.
\medskip
 
\begin{keywords}
pneumatic artificial muscle; angle-and-stiffness control; sensor-less control; unscented kalman filter
\end{keywords}\medskip

\end{abstract}

\section{Introduction}

The McKibben pneumatic artificial muscle (PAM) is an actuator that provides a high force-to-weight ratio, and it is lightweight and has excellent flexibility.
These properties enable the construction of a lightweight, highly backdrivable, and direct-drive actuator. 
Therefore, a PAM is a suitable actuator for devices that often contact human beings, such as assistant robots, nursing care robots, and rehabilitation orthoses~\cite{2011_Survey}.
The PAM comprises an internal rubber bladder surrounded by inextensible threads that are braided in a spiral.
When compressed air is supplied to a PAM, the diameter of the rubber bladder increases, causing the long axis of the PAM to contract.
Though the PAM generates a contraction force when compressed air is supplied, it does not generate any force when the compressed air is released.
An antagonistic structure consisting of two PAMs arranged in parallel, with one PAM connected to the other via a joint, is often used to realize rotational motion and perform behaviors similar to those of human muscles~\cite{10_Mihn, 2011_Beyl, 2012_Sui, 13_Andri,14_Andri, 2016_Andrikopoulos, 20_Jamwal}.
However, the PAM exhibits high nonlinearity owing to its pressure dynamics, valve characteristics, and friction, making its modeling and control quite challenging.

The PAM is a variable-stiffness actuator with adjustable rigidity or hardness.
Variable stiffness can enhance contact and collision safety between robots and humans or environments and improve the comfort of wearing rehabilitation orthotics. 
In the case of rehabilitation, the stiffness prescribed by physiotherapists changes according to the treatment phase; thus, it must be adjusted step-by-step. 
There have been several studies on a model-based stiffness or compliance control using a PAM actuator~\cite{2009_nakamura, 2010_stiffness, 2011_Choi, 2012_Sui, 2015_Ugurlu, 2015_Saito, 2017_okajima, 2017_Sugimoto, 2018_Cao, 2017_Zhao, 2019_Ugurl}, and some of the studies are as follows.
Cao \textit{et al.}~\cite{2018_Cao} proposed model-based angle-compliance control to develop a robotic gait rehabilitation device.
Ugurlu \textit{et al.}~\cite{2019_Ugurl} realized simultaneous control of position and stiffness in an antagonistically driven PAM actuator.
These studies experimentally determined a reference in joint stiffness, while they did not clarify an admissible reference set for joint stiffness and angle.
Clarification of the admissible information is essential for the practical implementation and systematic design of PAM actuators for applications such as antagonistic force maximization~\cite{2017_Dirven}.

The use of sensors such as an encoder and a force sensor can provide efficient tracking performance of a PAM actuator. 
However, a force sensor is relatively expensive and heavy. 
As a result, a sensor-less approach have been proposed.
For example, a static force map was used in~\cite{2015_polymodel} to estimate the joint torque of a PAM actuator.
A static force map was also used in~\cite{sensorless_PAM} to control an antagonistic PAM joint actuator system.
If the joint angle can be estimated, the cost of designing and producing a PAM actuator could be further reduced.
Our previous study of~\cite{APAMmodel} proposed a detailed mathematical model of an antagonistic PAM actuator system.
The feature of the study is that a pressure sensor is used to estimate PAM's joint angle and torque.
An accurate PAM model is likely to be helpful for estimating joint stiffness as well as angle because both are related to the working pressure.

The purpose of this paper, therefore, is to propose the simultaneous sensor-less control method for the joint angle and stiffness of an antagonistic PAM actuator system, and to develop a procedure to obtain a set of admissible references, defined as pairs of stiffnesses and joint angles.
To realize sensor-less control using only pressure measurements, 
this study uses an unscented Kalman filter (UKF) to estimate the joint angle and contraction forces by employing the detailed model in~\cite{APAMmodel}.
The contribution of this study is that the proposed sensor-less angle/stiffness control method using a UKF represents a novel approach in robotics.  
Moreover, it  does not require any encoder, which previous relevant studies \cite{2009_nakamura, 2010_stiffness, 2011_Choi, 2012_Sui, 2015_Ugurlu, 2015_Saito, 2017_okajima, 2017_Sugimoto,  2017_Zhao, 2018_Cao, 2019_Ugurl} have relied upon, to achieve simultaneous control. 
This sensor-less angle/stiffness control approach can realize a low-cost, lightweight actuator that ensures safe contact with humans and environments.
Furthermore, this paper also presents experimental results of the proposed control method using a previously developed antagonistic PAM actuator system testbed.
The results indicate that the reference admissible set obtained using the model is useful in choosing allowable references and confirm that the reference set covers the experimentally obtained trajectories for the stiffness and angle. 
Three scenarios are chosen from the characterized reference admissible set, and it is confirmed that the proposed method can independently control the stiffness and angle.

The remainder of this paper is organized as follows: 
Section \ref{sec:apam} presents the antagonistic PAM actuator system and its mathematical model and derives the expressions for the joint stiffness.
Section \ref{sec:system} proposes the angle/stiffness control system and describes its details.
Section \ref{sec:cntrl} presents the experimental results of sensor-less angle/stiffness control.
Finally, Section \ref{sec:conc} concludes this paper. 

\section{Antagonistic PAM Actuator System and Its Joint Stiffness Model}\label{sec:apam}
This section briefly introduces a practical antagonistic PAM actuator system and its mathematical model, as presented in our previous study \cite{APAMmodel}, 
and then mathematically describes the joint stiffness using this model.

\subsection{Experimental Setup of PAM Actuator System}
\begin{figure}[t]
	\centering
	\subfigure[hotograph of antagonistic PAM joint actuator.]{
	\includegraphics[width=0.55\hsize]{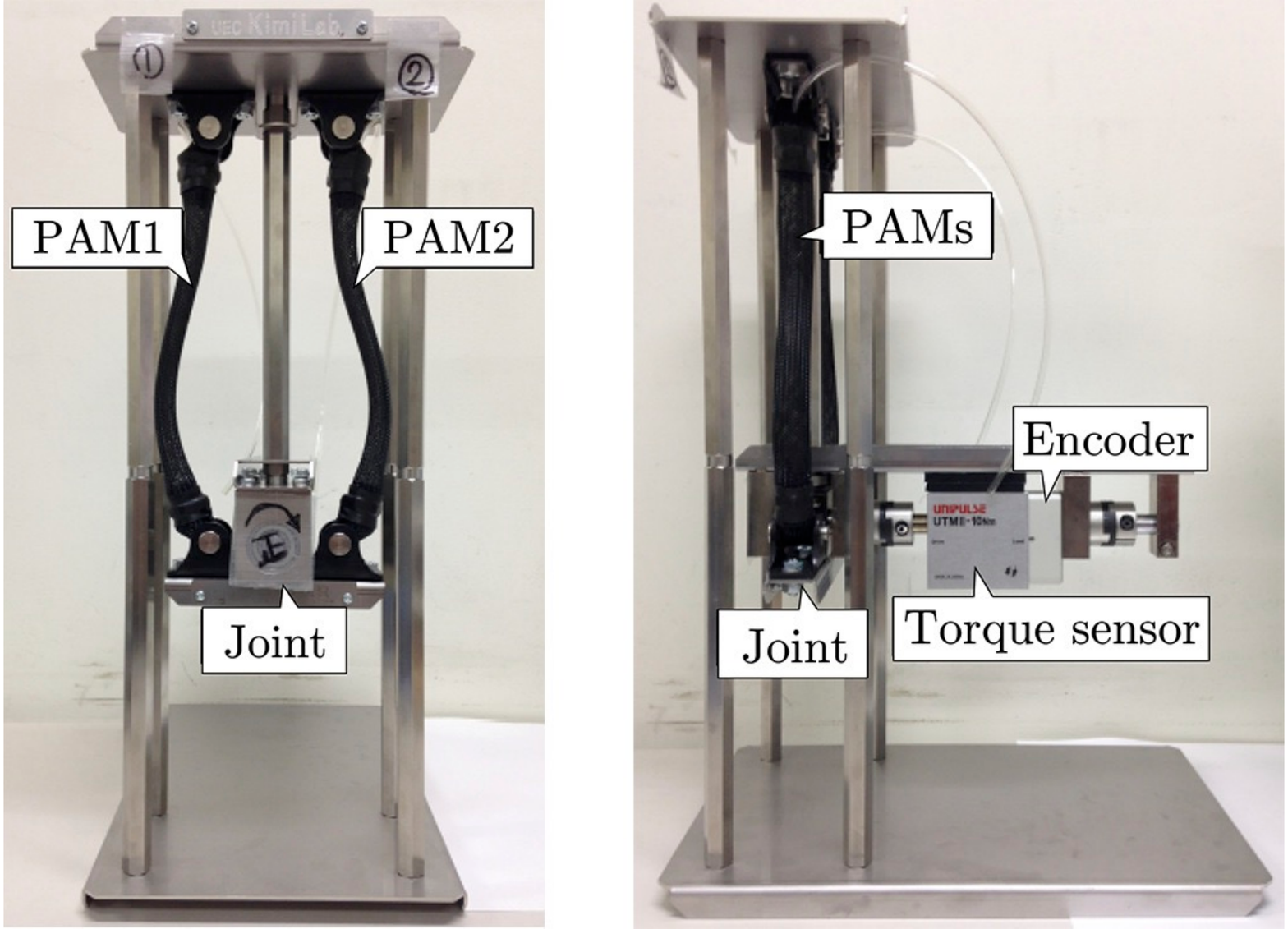}
	\label{sfig:atgsys}}
	\subfigure[Schematic of antagonistic PAM system.]{
	\includegraphics[width=0.55\hsize]{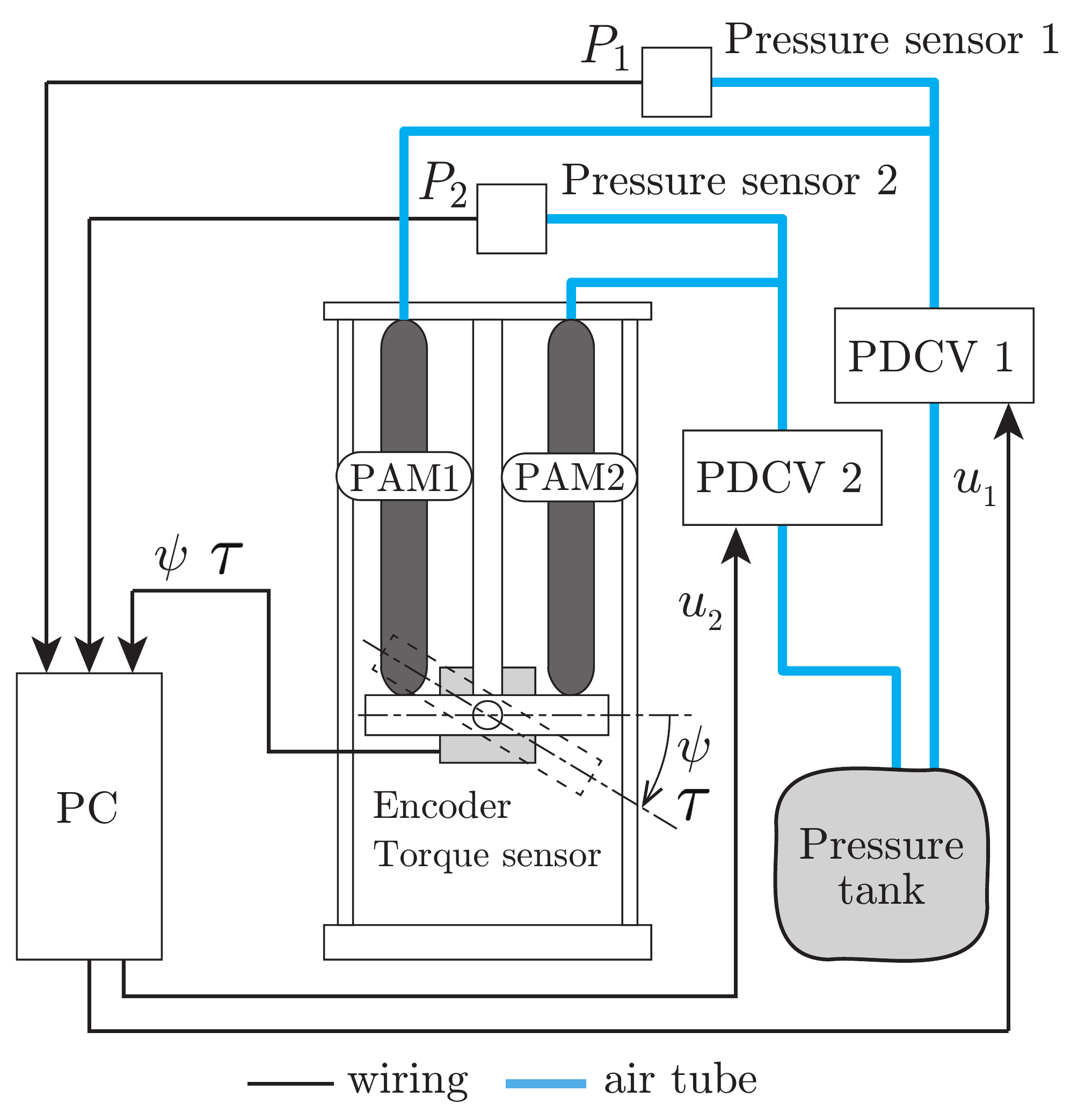}
	\label{sfig:eqatg}}
	\caption{Antagonistic PAM system \cite{APAMmodel}.}
	\label{fig:setup}
\end{figure}

Figures. \ref{fig:setup}\subref{sfig:atgsys} and \subref{sfig:eqatg} respectively depict a photograph and schematic of the antagonistic PAM actuator system used in this study.
The system consists of two PAMs~(Airmusle, Kanda Tsushin Kogyo) connected to each other by a seesaw-like joint part, two proportional directional control valves (PDCVs) (5/3-way valve, FESTO), a pressure tank~(6-25, JUN-AIR), two pressure sensors~(E8F2-B10C, OMRON), a torque sensor with an encoder (TM II-10 Nm(R), UNIPULSE), and a PC.
The length and diameter of the PAMs are respectively 170~mm and 12.7~mm.
The pressure tank stores compressed air and is connected to the PDCVs and PAMs by air tubes.
Air flows adjusted by the PDCVs are used to drive the PAMs and rotates the joint part.
The encoder and torque meter respectively measure the joint angle and torque, and the pressure sensors measure the inner pressure of the PAMs.
The voltage signals to the two PDCVs ($u_1$ and $u_2$) are the inputs of the system, the joint angle $\psi$, inner pressure of the PAMs $(P_1$ and $P_2$), torque $\tau$, and joint stiffness $K_P$ are the outputs of the system.
The values of $\psi,\ P_1,\ P_2,$ and $\tau$ are obtained by the sensors, and $K_P$ is obtained using the equation derived below.
The PC has a 3.2~GHz CPU and 8~GB of RAM; the operating system is Ubuntu~12.04 and the the preemption-patched Xenomai~2.6.2.1 is installed.
The sampling period, $T_{\rm stp}$, was set to 1 ms. 
The range of the joint angle is $\pm$ 25 deg, and the range of the output torque is $\pm$ 3.0 Nm.

\subsection{Brief Introduction of PAM Actuator Model}
A state-space model that considers the noises in the antagonistic PAM system is given as follows:
\begin{subequations}
	\begin{align}
		\dot{x}(t) &= f_\sigma(x(t), u(t)) + v(t) \hspace*{3ex}{\sf if} \ \ x(t) \in \mathcal{X}_{\sigma}, \label{eq:pam_a} \\
		y(t) &= h(x(t)) + w(t), \label{eq:pam_b}
	\end{align}
	\label{eq:pam}
\end{subequations}
\!\!where $t \in \mathbb{R}_{\geq0}$ is the time; 
$\mathbb{R}_{\geq0}$ is a nonnegative real number set;
$u:=[u_1 \ u_2]^\mathrm{T}\in{\mathcal U}\subset \mathbb{R}^2$ is the input voltages to the two PDCVs, 
in which ${\mathcal U}:=[0,10]^2$ is a set of allowable inputs determined by the characteristics of the PDCV;
the state variable is $x:=[\psi \ \dot{\psi} \ P_1 \ P_2 ]^\mathrm{T} \in \mathbb{R}^4$, and the output variable is $y:=[\psi \ P_1 \ P_2 \ \tau\ K_P]^\mathrm{T} \in \mathbb{R}^5$, in which ${\mathcal P}:=[200,750]$ is defined as a set of allowable pressures determined by the specification of the PAMs, i.e., $P_1, P_2\in\mathcal{P}$; 
$v$ and $w$ are the process noise and observation noise, respectively;
$f_{\sigma}:\mathbb{R}^4\rightarrow \mathbb{R}^4$ is a nonlinear function with 18 subsystems, and it switches according to if-then rules; 
${\mathcal X}_\sigma: =\{x \in \mathbb{R}^4 | \Psi_{\sigma}(x)>0 \}$ is the state set, where $\sigma\in \Sigma:=\{1, 2, \cdots, 18\}$ is the subsystem's index;
$\Psi_{\sigma}(x)$ is a function derived from the modes in the form of if-then rules; 
finally, the function $h:\mathbb{R}^4\rightarrow \mathbb{R}^5$ is an observation equation.
Tables~\ref{tbl:variables} and \ref{tbl:parameters2pam} respectively show the model parameters and their identified values for the antagonistic PAM actuator.
Details and parameter settings of the PAM actuator model can be found in \cite{APAMmodel}.

\subsection{Joint Stiffness}\label{subsec:stiff}
The joint stiffness equation describing the antagonistic PAM actuator system is based on the derivation process, discussed in \cite{2019_Ugurl}.
However, this study derives a different-form of equation for joint stiffness because the mechanical structures of the actuator and PAM force model differ.

\begin{figure}[t]
	\centering
	\includegraphics[scale = 0.65]{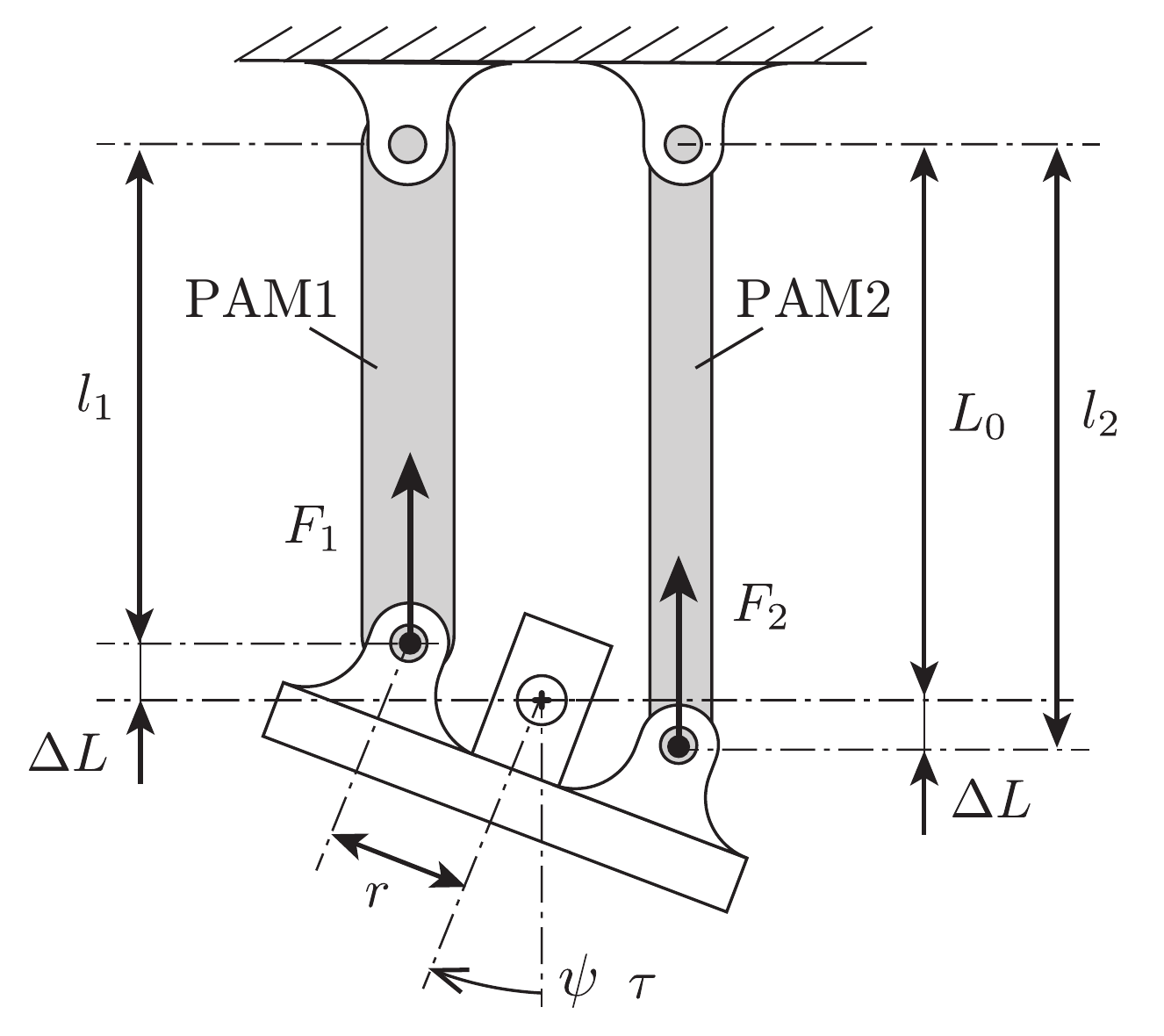}\\
	\caption{Geometric structure of antagonistic PAM joint.}
	\label{fig:apam}
\end{figure}

Considering the geometric relationship shown in Figure.\ \ref{fig:apam}, the lengths of the two PAMs, $l_1$ and $l_2$, are respectively given by
\begin{align}
	l_1(t)=L_0-\Delta L (t) ,\ l_2(t)=L_0+\Delta L (t), \label{eq:l}
\end{align}
where the horizontal displacement of the two PAMs due to rotation is so small that it can be neglected, that is, $\Delta L (t) \approx r \sin \psi(t)$; $r$ is the radius of the seesaw; and $L_0$ is the PAM length when the seesaw is at the horizontal position. 

The PAM contraction force has static characteristics that relate inner pressure under the joint fixed \cite{2011_Minh}.
The contraction force $F$ can therefore be described as a function of the PAM inner pressure and length as follows:
\begin{align}
	F_i(l_i(t),P_i(t))=v_i(l_i(t)) P_i(t)+w_i(l_i(t)), 
	\label{eq:force}
\end{align}
where $i\in{\mathcal I}:=\{1,2\}$ and 
\begin{subequations}
	\begin{align}
		v_i(l_i(t)) &= p_{v_{1i}} l_i(t) + p_{v_{2i}},\label{eq:pamfrc1}\\
		w_i(l_i(t)) &= p_{w_{1i}} l_i(t) + p_{w_{2i}}.\label{eq:pamfrc2}
	\end{align}
	\label{eq:pamfrc}
\end{subequations}
\!\!The joint torque $\tau$ can be described as follows:
	\begin{align}
		\tau(t) = r\cos \psi(t) \left( F_1(t) - F_2(t) \right).
		\label{eq:tauPAM}
	\end{align}
Joint stiffness is used in this study as an index to describe the joint resistance to an applied moment.
As the joint stiffness decreases, the joint becomes more flexible in response to external forces; thus, the interaction between robots and people or environments becomes safer.
The joint stiffness, denoted as $K_{P}$, is defined as the partial differentiation of the joint torque \eqref{eq:tauPAM} as follows:
	\begin{align}
		K_{P}(t)=& -\frac{\partial \tau(t)}{\partial \psi(t)} \nonumber \\
		=&\ r\sin\psi(t) (F_1(t) - F_2(t)) \nonumber \\
		&\ -r\cos\psi(t) \left(\frac{\partial F_1(t)}{\partial l_1(t)}\frac{\partial l_1(t)}{\partial \psi(t)}-\frac{\partial F_2(t)}{\partial \psi(t)}\frac{\partial l_2(t)}{\partial \psi(t)}\right) \nonumber \\
		=&\ r\sin\psi(t) (F_1(t) - F_2(t)) \nonumber \\
		&\ + r^2\cos^2\psi(t) \left(\frac{\partial F_1(t)}{\partial l_1(t)} + \frac{\partial F_2(t)}{\partial l_2(t)} \right).\label{eq:KP}
	\end{align}
Applying \eqref{eq:pamfrc} to \eqref{eq:force}, the PAM contraction forces and the partial differentiation of the PAM length can be obtained:
	\begin{align*}
		F_i(t) &= (p_{v1i}l_i(t) + p_{v2i})P_i(t) + (p_{w1i}l_i(t) + p_{w2i}),\\
		\frac{\partial F_i(t)}{\partial l_i(t)} &= p_{v1i}P_i(t) + (p_{v1i}l_i(t) + p_{v2i})\frac{\partial P_i(t)}{\partial l_i(t)} + p_{w1i}. \nonumber
	\end{align*}
Note that $\partial P_i/\partial l_i$ is negligible when the volume of the PAM is much smaller than that of the air tank\cite{2019_Ugurl};  asa result, \eqref{eq:KP} can be written as follows:
	\begin{align*}
		K_{P}&=\ r\sin\psi(t) (F_1(t) - F_2(t)) \\
		&\ +r^2\cos^2\psi(t) ( p_{v11}P_1(t) +p_{w11} + p_{v12}P_2(t) +p_{w12}).
		\end{align*}
By defining $\alpha_i(t) := p_{v2i}P_i(t) + p_{w2i} $, $K_P$ can be described as
	\begin{align}
		K_{P}&=\ r\sin\psi(t) (F_1(t) - F_2(t)) \label{eq:K} \\
		&\ + r^2\cos^2\psi(t) \left( \frac{F_1(t) - \alpha_1(t)}{l_1(t)} + \frac{F_2(t) - \alpha_2(t)}{l_2(t)} \right).  \notag
	\end{align}
	
\begin{table}[tb]
	\tbl{Parameters of Antagonistic PAM system}{
	\begin{tabular}[t]{llc} 
		\hline
		\hline
		$r_p$ & : radius of shaft (m) & \ \\
		$r$ & : radius of seesaw (m) &\ \\
		$L_0$ & : initial length of PAM (m) &\ \\
		$M$ & : weight of seesaw (kg) &\ \\
		$g$ & : gravitational acceleration (m/s$^2$) &\ \\
		$P_{\rm tank}$ & : source absolute pressure (Pa) &\ \\
		$P_{\rm out}$ & : atmospheric pressure  (Pa) &\ \\
		$k$ & : specific heat ratio for air (--) &\ \\
		$R$ & : ideal gas constant (J/kg$\cdot$K) &\ \\
		$T$ & : absolute temperature (K) &\ \\
		$J$ & : moment of inertia of seesaw (kg$\cdot$m$^2$) &\ \\
		$k_s$ & : coefficient of static torque of seesaw (N$\cdot$m/rad) &\ \\
		$c_s$ & : viscous friction coefficient (N$\cdot$s) &\
		\raisebox{6.5mm}[0pt][0pt]{\hbox{\rotatebox{90}{Directly measurable}}}\\
		\hline
		$D_1,  D_2,  D_3$ & : coefficients of polynomial (m, m$^2$, m$^3$) &\ \\
		$p_{v1i}$, $p_{v2i}$, $p_{w1i}$, $p_{w2i}$  & : coefficient of force for PAM$i$ (--) &\ \\ 
		$A_{1i}$, $A_{2i}$ & : orifice area of PDCV$i$ (m$^2$) &\ \\
		$k_1,  k_2$ & : polytropic indexes (--) &\ \\
		$T_p'$ & : Coulomb friction coefficient of PAM (--) &\ \\
		$\mu_s$ & : Coulomb friction coefficient of shaft (--) &\
		\raisebox{3.0mm}[0pt][0pt]{\hbox{\rotatebox{90}{Estimated}}}\\
		\hline
		\hline
	\end{tabular}}
	\label{tbl:variables}
\end{table}

	\begin{table}[t]
		\tbl{Identified parameters of antagonistic PAM system}{
		\begin{tabular}{lr | lr}
		\hline
		\hline
		parameter & value & parameter & value\\
		\hline
		$r_p$ (m) & 0.006 & $D_1$ (m) & $-2.440 \times 10^{-2}$\\
		$r$ (m) & 0.0365 & $D_2$ (m$^2$) & $6.824 \times 10^{-3}$\\
		$L_0$ (m) & $0.165$ & $D_3$ (m$^3$) & $-4.254 \times 10^{-4}$\\
		$M$ (kg) & $0.256$ & $p_{v11}$ (--)& $7.045 \times 10^{-3}$\\
		$g$ (m/s$^2$) & $9.80$ & $p_{v21}$ (--)& $-1.017 \times 10^{-3}$\\
		$P_{\rm tank}$ (Pa) & $0.7100\times 10^{6}$ & $p_{w11}$ (--)& $-5.568 \times 10^{2}$\\
		$P_{\rm out}$ (Pa) & $0.1013 \times 10^{6}$ &$p_{w21}$ (--)& $72.86$\\
		$k$ (--) & $1.40$ & $p_{v12}$ (--)& $6.423 \times 10^{-3}$\\
		$R$ (J/kg$\cdot$K) & $287$ & $p_{v22}$ (--)& $-9.184 \times 10^{-4}$\\
		$T$ (K) & $293$ & $p_{w12}$ (--)& $-197.8$\\
		$J$ (kg$\cdot$m$^2$) & $4.263 \times 10^{-4}$ & $p_{w22}$ (--)& $-15.75$\\
		$k_s$ (N$\cdot$m/rad) & $4.117 \times 10^{-4}$ & $A_{11}$ (m$^2$) & 5.184 $\times 10^{-8}$\\
		$c_s$ (N$\cdot$s) & $2.256 \times 10^{-3}$ & $A_{12}$ (m$^2$) & 7.776 $\times 10^{-8}$\\
		$k_1$ (--) & 1.100 & $T_p'$ (--) & $4 \times 10^{8}$\\
		$k_2$ (--) & 0.4545 & $ \mu_s$ (--) & 0.2\\
		\hline
		\hline
		\end{tabular}}
		\label{tbl:parameters2pam}
	\end{table}

\section{Design of Sensor-less Control System with Admissible References}\label{sec:system}
This section describes the proposed simultaneous sensor-less control method for the angle and stiffness and discusses a procedure for depicting a reference admissible set on an angle--stiffness plane.

\subsection{Sensor-less Angle/Stiffness Control}
\begin{figure*}[t]
	\centering
	\includegraphics[width=0.85\hsize]{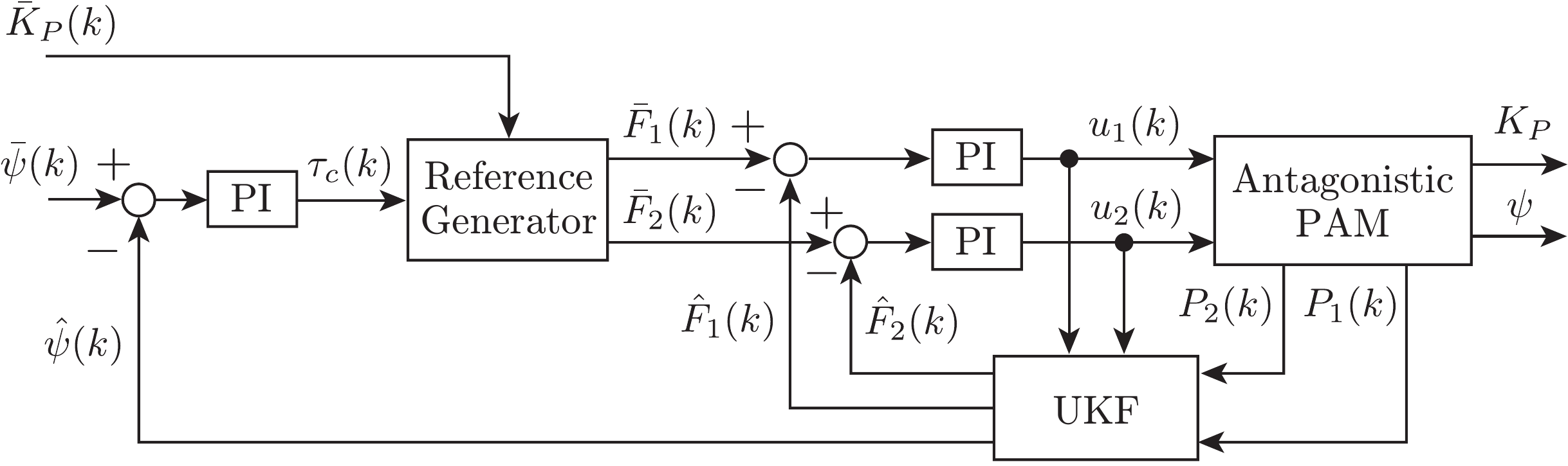}\\
	\caption{Block diagram of sensor-less angle/stiffness control system.}
	\label{fig:AK_controller}
\end{figure*}

The sensor-less angle/stiffness control system proposed in this study consists of a reference generator, PI controllers, and a UKF.
A block diagram of the proposed control system is shown in Figure.\ \ref{fig:AK_controller}; a similar control system configuration was applied in \cite{2019_Ugurl} using force sensors for feedback control.
The detailed structure of the control system is described below.

\subsubsection{Reference generator}
The reference generator provides adequate contraction force reference signals to each of the PAMs ($\bar{F}_1$ and $\bar{F}_2$) using a computed torque command $\tau_c$ and a given reference joint stiffness $\bar{K}_P$.
The relation among $\bar{F}_1$, $\bar{F}_2$, $\tau_c$, and $\bar{K}_P$ can be derived algebraically from the obtained stiffness equation.
Using $\bar{F}_1$, $\bar{F}_2$, $\tau_c$, and $\bar{K}_P$, \eqref{eq:tauPAM}, and \eqref{eq:K} can be rewritten as follows:
\begin{align*}
	\tau_c(t) = r\cos \psi(t) \left( \bar{F}_1(t) - \bar{F}_2(t) \right),
\end{align*}
\begin{align*}
	\bar{K}_P &= \ r\sin\psi(t) (\bar{F}_1(t) - \bar{F}_2(t)) \nonumber \\
	&+ r^2\cos^2\psi(t) \left( \frac{\bar{F}_1(t) - \alpha_1(t)}{l_1(t)} + \frac{\bar{F}_2(t) - \alpha_2(t)}{l_2(t)} \right).
\end{align*}

Then, $\bar{F}_1$ and $\bar{F}_2$ can be obtained from the above two equations as follows:
\begin{align}
	\bar{F}_{1}(t) = &\ \frac{1}{r^2\cos^2\psi(t)}\frac{l_1(t)l_2(t)}{l_1(t) + l_2(t)}\biggl[ \bar{K}_P(t) \nonumber \biggr.\\
	&\ \left.+ \Bigl(\frac{r\cos\psi(t)}{l_2(t)} - \tan(\psi(t))\Bigr)\tau_c\right.\nonumber \\
	&\ \biggl.- r^2\cos^2\psi(t)\Bigl(\frac{\alpha_1}{l_1(t)} + \frac{\alpha_2}{l_2(t)}\Bigr) \biggr], \label{eq:F1_generate}\\
	\bar{F}_{2}(t) = &\ \bar{F}_{	1}(t) - \frac{\tau_c}{r\cos\psi(t)}.
	\label{eq:F2_generate}
\end{align}
The error between the reference angle $\bar{\psi}$ and the measured angle $\hat{\psi}$ is then converted into a command torque $\tau_{c}$ by the feedback PI controller according to:
\begin{align*}
	\begin{cases}
		x^\psi(k+1) = x^\psi(k) + T_{\rm stp}e^\psi(k),\\
	    \tau_c(k) = G_{\rm I}^\psi x^\psi(k) + G_{\rm P}^\psi e^\psi(k),
	\end{cases}
\end{align*}
where $e^\psi := \bar{\psi} - \hat{\psi}$; 
$x^\psi\in\mathbb{R}$ is the controller state; 
$G_{\rm P}^\psi$ and $G_{\rm I}^\psi$ are respectively the proportional and integral gains, which were set to 15 and 10, respectively.
Based on the command torque $\tau_c$ and reference joint stiffness $\bar{K}_P$, force signals ($\bar{F}_1$ and $\bar{F}_2$) are generated by the reference generator according to \eqref{eq:F1_generate} and \eqref{eq:F2_generate}, respectively.
The resulting values of $\bar{F}_1$ and $\bar{F}_2$ are then compared with the estimated forces $\hat{F_1}$ and $\hat{F_2}$, respectively, and the resulting errors are fed to the PI controllers to generate control input voltages $u_1$ and $u_2$, respectively.
These controllers are governed as follows:
\begin{align*}
	\begin{cases}
		x_i^{F}(k+1) = x_i^F(k) + T_{\rm stp}e_i^F(k),\\
	    u_i(k) = G_{\rm I}^Fx_i^F(k) + G_{\rm P}^F e_i^F(k),
	\end{cases}
\end{align*}
where $i\in{\mathcal I}$, $e_i^F := \bar{F_i} - \hat{F_i}$ and $x_i^F\in\mathbb{R}$ is the controller state.
The proportional gain $G_{\rm P}^F$ and the integral gain $G_{\rm I}^F$ of the force controllers were respectively set to 0.08 and 0.15 using trial and error.

\subsubsection{State estimator}\label{ssubsec:UKF}
This study uses the UKF \cite{ukf_2000, ukf_2015} to estimate the state of the antagonistic PAM actuator system based on the pressure sensor information; the rotary encoder is used only to evaluate the resulting estimation.
The UKF applies the pressure sensor information and the actuator model from \cite{APAMmodel} to estimate the joint angle and torque, and the torque is used to compute the stiffness.

A discrete-time representation of the actuator model \eqref{eq:pam} used in the UKF can be given as follows by the fourth-order Runge-Kutta method:
\begin{subequations}
	\begin{align*}
		x(k+1) &= f_\sigma(x(k), u(k)) + v(k) \hspace*{1ex}{\sf if} \ x(k) \in \mathcal{X}_{\sigma},\hspace*{2ex} \\
		y(k) &= g(x(k)) + w(k),
	\end{align*}
	\label{eq:pam_ukf}
\end{subequations}
\!\!with
\begin{align*}
	g(x(k)) = 
	\begin{bmatrix}
	0 & 0 & 1 & 0\\
	0 & 0 & 0 & 1
	\end{bmatrix}x(k),
\end{align*}
where $v\in \mathbb{R}^4$ and $w\in \mathbb{R}^2$ are the process and observation noise of the system, respectively;
$v$ and $w$ are zero-man white noises with the covariance matrices $Q\in \mathbb{R}^4\times \mathbb{R}^4$ and $R\in \mathbb{R}^2 \times \mathbb{R}^2$, respectively.
Table~\ref{tbl:covariance} lists the UKF's parameters applied in this study.
The estimated PAM forces ($\hat{F}_1$ and $\hat{F}_2$) are obtained by substituting the estimated state variables into \eqref{eq:force}.

\begin{table}[t]
	\tbl{parameters of Unscented Kalman Filter.}{
	\renewcommand{\arraystretch}{1.5}
		\begin{tabular}{lrlr}
		\hline
		\hline
		parameter & value \\
		\hline
		$\bm{P}(0)$ & ${\rm diag}(10^{-5},10^{-4},\ 10^{6},\ 10^{6})$ \\
		$Q$	& ${\rm diag}(10^{-5},10^{-4},\ 10^{6},\ 10^{6})$ \\
		$R$ & ${\rm diag}(10^8,10^8)$\\
		$\kappa$ & 0\\
		\hline
		\hline
		\end{tabular}}
	\label{tbl:covariance}
\end{table}

\subsection{Reference Admissible Set}\label{subsec:refset}
When considering a tracking control problem, prior information of an allowable reference helps an operator to set reference values.
Thus, this study introduces a reference admissible set $W\subset\mathbb{R}^2$, which is defined as follows,
\begin{align*}
W:=\left\{ [\,\psi\ \,K_P\,]^T\in\mathbb{R}^{2}\,|\,\right. & \left. \forall P_i(\infty)\in\mathcal{P}, \ \forall i\in\mathcal{I}, \right.\\
&\left. \exists \kappa_u(P_1(\infty), P_2(\infty))\in\mathcal{U}, \right.\\
& \left. 0=f_\sigma(x(\infty),\kappa_u(P_1(\infty),P_2(\infty)), \right.\\
& \left. y(\infty)=h(x(\infty)) \right\},
\end{align*}
where $x(\infty)$ and $y(\infty)$ denote a state and corresponding output of the antagonistic PAM system \eqref{eq:pam} in a steady state, 
and $\kappa_u$ is a function that yields control inputs corresponding to pressures $P_i$ in a steady state.
This reference admissible set can be computed as explained below.

\begin{figure}[t]
	\centering
	\subfigure[Static relationships between two PAM pressures and joint angles.]
	{\includegraphics[width=0.5\hsize]{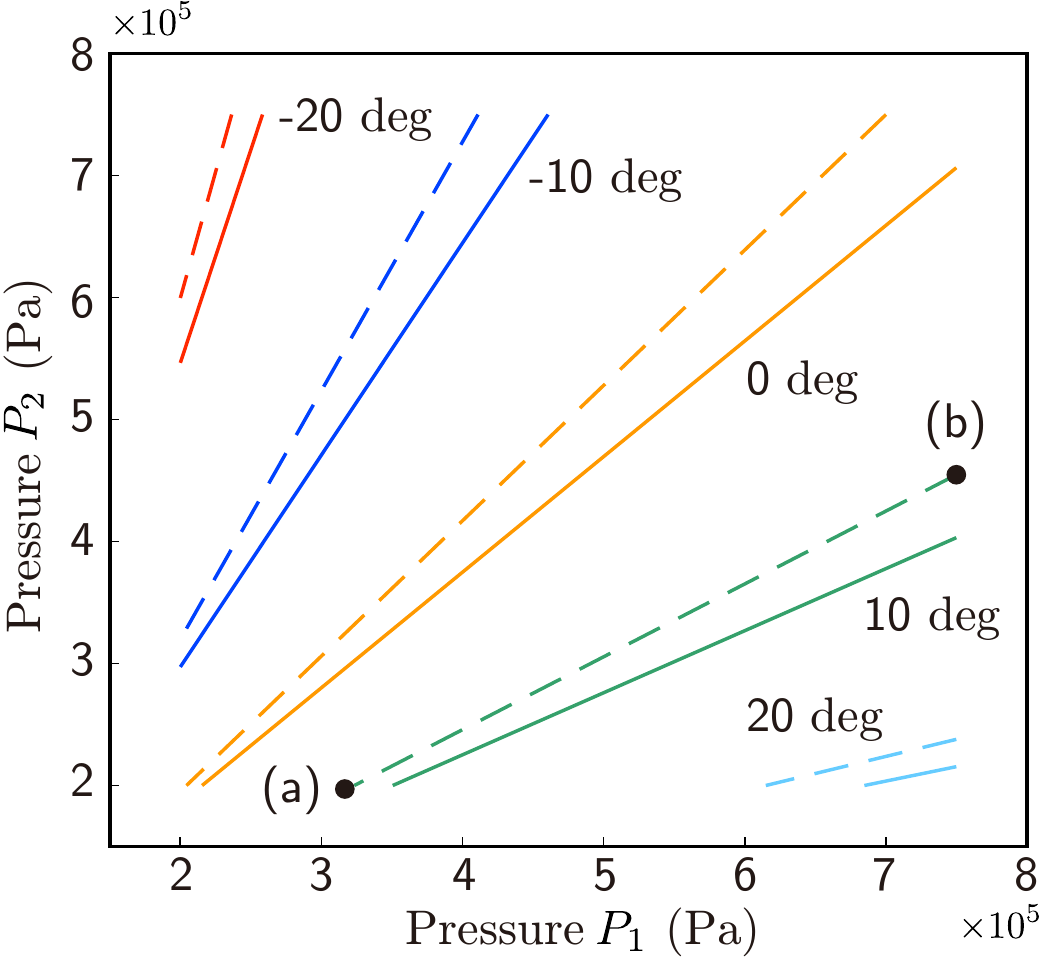}
	\label{sfig:steady}}\\
	\subfigure[Reference admissible set.]
	{\includegraphics[width=0.5\hsize]{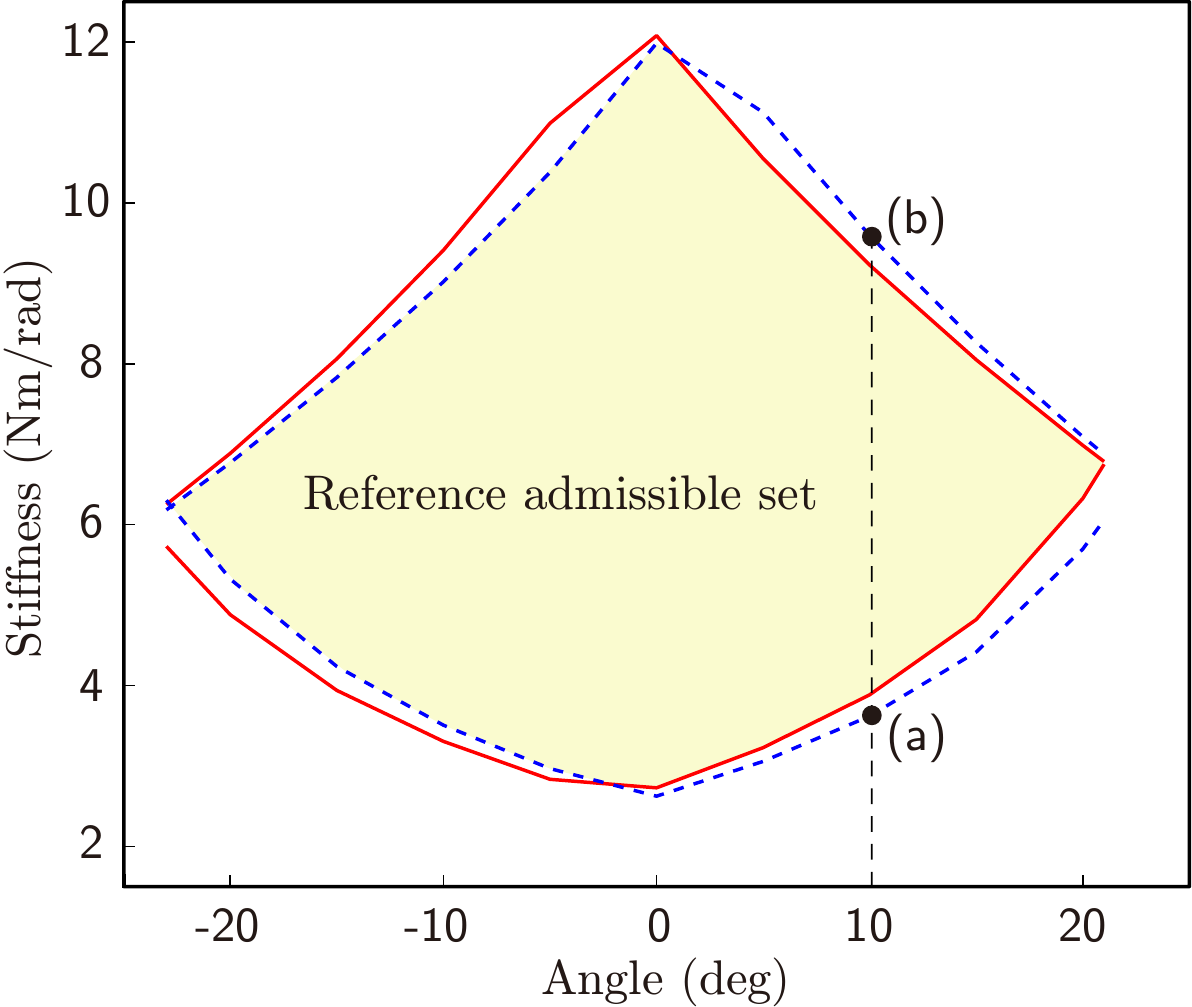}
	\label{sfig:region}}
	\caption{Computation of a reference admissible set using the presented procedure.}	
	\label{fig:region}
\end{figure}

In a steady state, the derivative term in the equation of the seesaw motion~(Eq.~(8) in \cite{APAMmodel}) becomes zero so that the equation can be written as follows:
\begin{align}
	\psi(\infty)=\dfrac{\tau-T_f}{k_s}.
	\label{eq:st_psi}
\end{align}
where $T_f$ is the resistant torque due to friction, and it holds $T_f=T_0\ \land\ |T_0|\leq T_s+T_p$ in a steady state, in which $T_0$ is an external torque except for the frictional term, that is, $T_0 := \tau - k_s\psi$, and
\begin{align*}
	T_s &= r_p \mu_s |F_1(\infty) + F_2(\infty) - Mg|,\\
	T_p &= \mu_p\left( \frac{1}{(P_1(\infty) - P_{\rm out})^2}+\frac{1}{(P_2(\infty) - P_{\rm out})^2} \right).
\end{align*}
The friction torque $T_f$ takes a positive or negative value depending on the direction of the external torque.
Considering the maximum static friction, that is, $T_f=T_s+T_p$, \eqref{eq:st_psi} holds as follows:
\begin{subequations}
	\begin{align}
		\psi(\infty) &= \dfrac{\tau-(T_s+T_p)}{k_s}, \label{eq:psi1}\\
		\psi(\infty) &= \dfrac{\tau+(T_s+T_p)}{k_s}. \label{eq:psi2}
	\end{align}
	\label{eq:st_psi2}
\end{subequations}
\!\!By giving $P_1, P_2\in\mathcal{P}$ and using \eqref{eq:st_psi2}, the static relationship between the PAM inner pressures and the joint angle is illustrated in Figure.~\ref{fig:region}\subref{sfig:steady}, where the colored solid lines and dashed lines correspond to \eqref{eq:psi1} and \eqref{eq:psi2}, respectively.
As shown in the figure, different combinations of pressures correspond to achieve different joint angles.
Moreover, for a given particular joint angle, there exist certain combinations of pressures corresponding to the maximum and minimum stiffness.
For example, for a joint angle of 10 deg (the green-dashed line), the point marked in (a) corresponds to the minimum joint stiffness~($P_1=310$~(kPa) and $P_2=200$~(kPa)), and the point marked in (b) corresponds to the maximum joint stiffness~($P_1=750$~(kPa) and $P_2=450$~(kPa)). 
Using the points (a) and (b) enables the computation of the corresponding stiffness by \eqref{eq:K}, and these data are reported on an angle--stiffness plane in Figure.~\ref{fig:region}\subref{sfig:region}.
Finally, applying the same procedure to the operating joint angle range of the actuator, the reference admissible stiffnesses are represented by the light yellow-colored area in Figure.~\ref{fig:region}\subref{sfig:region}.
In this figure, the solid and dashed lines have the same meanings as those in Figure.~\ref{fig:region}\subref{sfig:steady}.
It should be noted that it is generally necessary to generate a pressure difference between two PAMs to obtain the desired angle; thus the smaller the amplitude of the joint angle is, the larger the range of the joint stiffnesses that can be set is.
In summary, the procedure applied in this study to compute a reference admissible set is described in {\bf Procedure~1}.

\begin{table}[t]
\centering
\begin{tabular}{llp{20em}}
\hline
\multicolumn{2}{p{7em}}{\textbf{Procedure 1}} & Computation of a reference admissible set~$W$. \\
\hline
$\textbf{Step 0:}$ & \multicolumn{2}{p{23em}}{Identify the model parameters listed in TABLE~\ref{tbl:variables}.}\\
$\textbf{Step 1:}$ & \multicolumn{2}{p{23em}}{Obtain the static relationships between the two pressures $P_1$, $P_2$, and the joint angle $\psi$ using (\ref{eq:psi1}) by giving $P_1, P_2\in\mathcal{P}$.}\\
$\textbf{Step 2:}$ & \multicolumn{2}{p{23em}}{Calculate two joint stiffness values corresponding to the maximum and minimum stiffness at a specific joint angle using \eqref{eq:K} and the static relationships obtained in {\bf Step 1}. }\\
$\textbf{Step 3:}$ & \multicolumn{2}{p{23em}}{Apply {\bf Step 2} to the entire driving angle range of the actuator system.}\\
$\textbf{Step 4:}$ & \multicolumn{2}{p{23em}}{Plot the relationships between the joint angle and stiffness obtained in  {\bf Steps 2} and {\bf 3} on an angle--stiffness plane.}\\
$\textbf{Step 5:}$ & \multicolumn{2}{p{23em}}{Apply the same procedure as in {\bf Steps 1} to {\bf 4} to (\ref{eq:psi2}).}\\
$\textbf{Step 6:}$ & \multicolumn{2}{p{23em}}{The area enclosed by both of the rhombic areas in the angle--stiffness plane obtained in {\bf Steps 4} and {\bf 5} is the reference admissible set~$W$.}\\
\hline
\end{tabular}
\end{table}

\section{Sensor-less Angle/Stiffness Control Experiments}\label{sec:cntrl}
This section demonstrates that the reference admissible set helps to determine a reference joint angle and stiffness, and the proposed control method is verified by conducting control experiments of the antagonistic PAM actuator system.

\subsection{Applicability of Reference Set}
\begin{figure}[t]
	\centering
	\subfigure[Time response of inner pressures of two PAMs.]
	{\includegraphics[width=0.55\hsize]{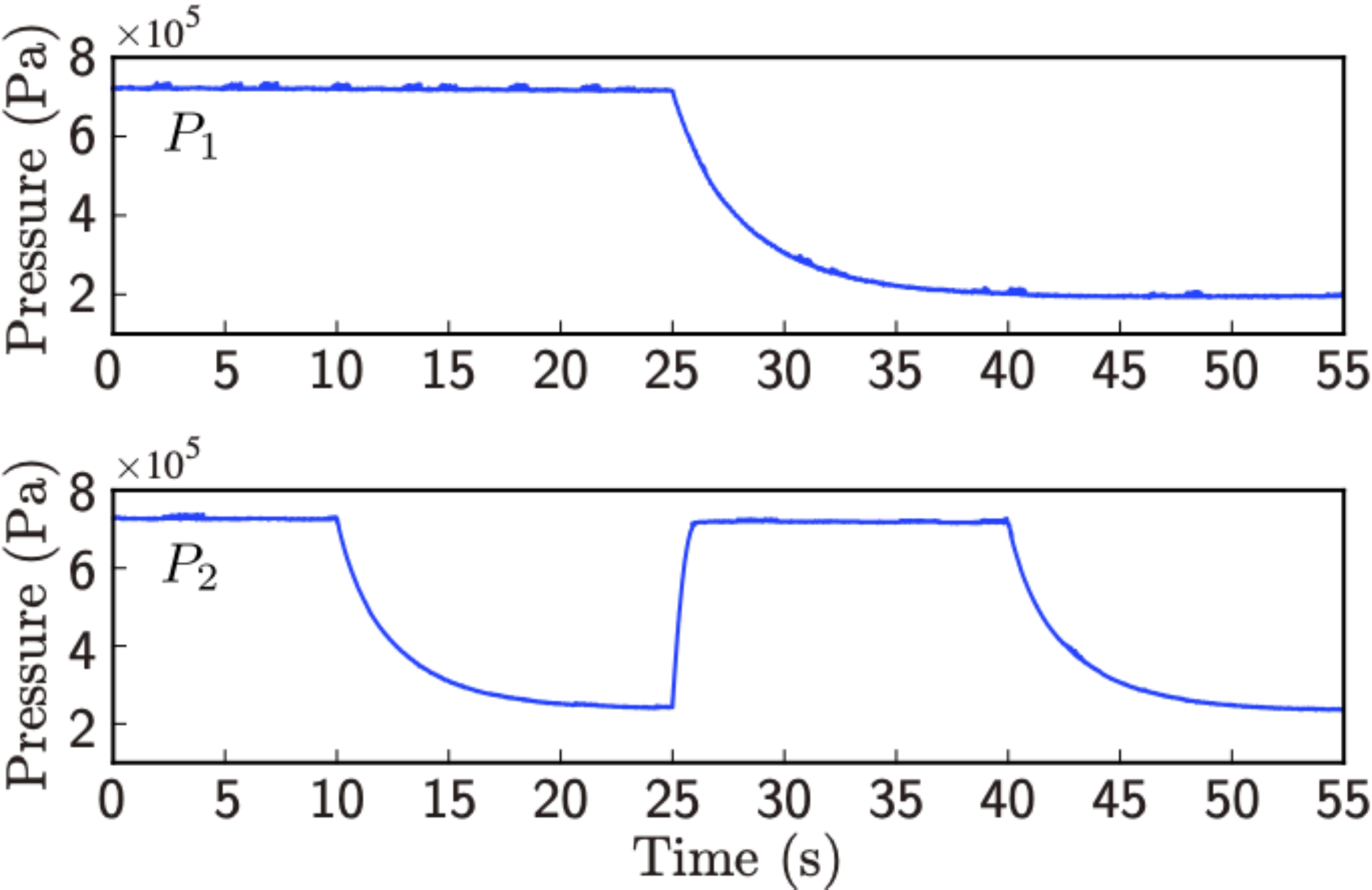}
	\label{sfig:pressure}}\\
	\subfigure[Time response of joint angle and stiffness.]
	{\includegraphics[width=0.55\hsize]{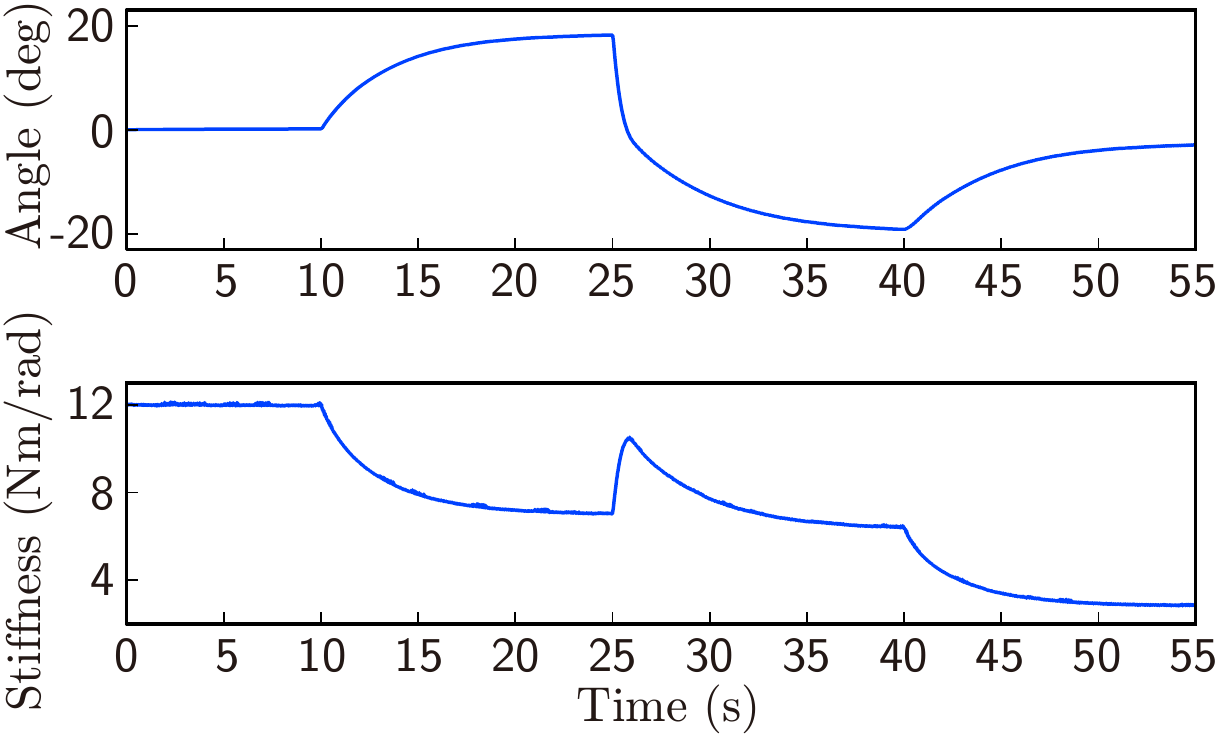}
	\label{sfig:ak}}\\
	\subfigure[Trajectyories on the computed reference admissible set.]
	{\includegraphics[width=0.5\hsize]{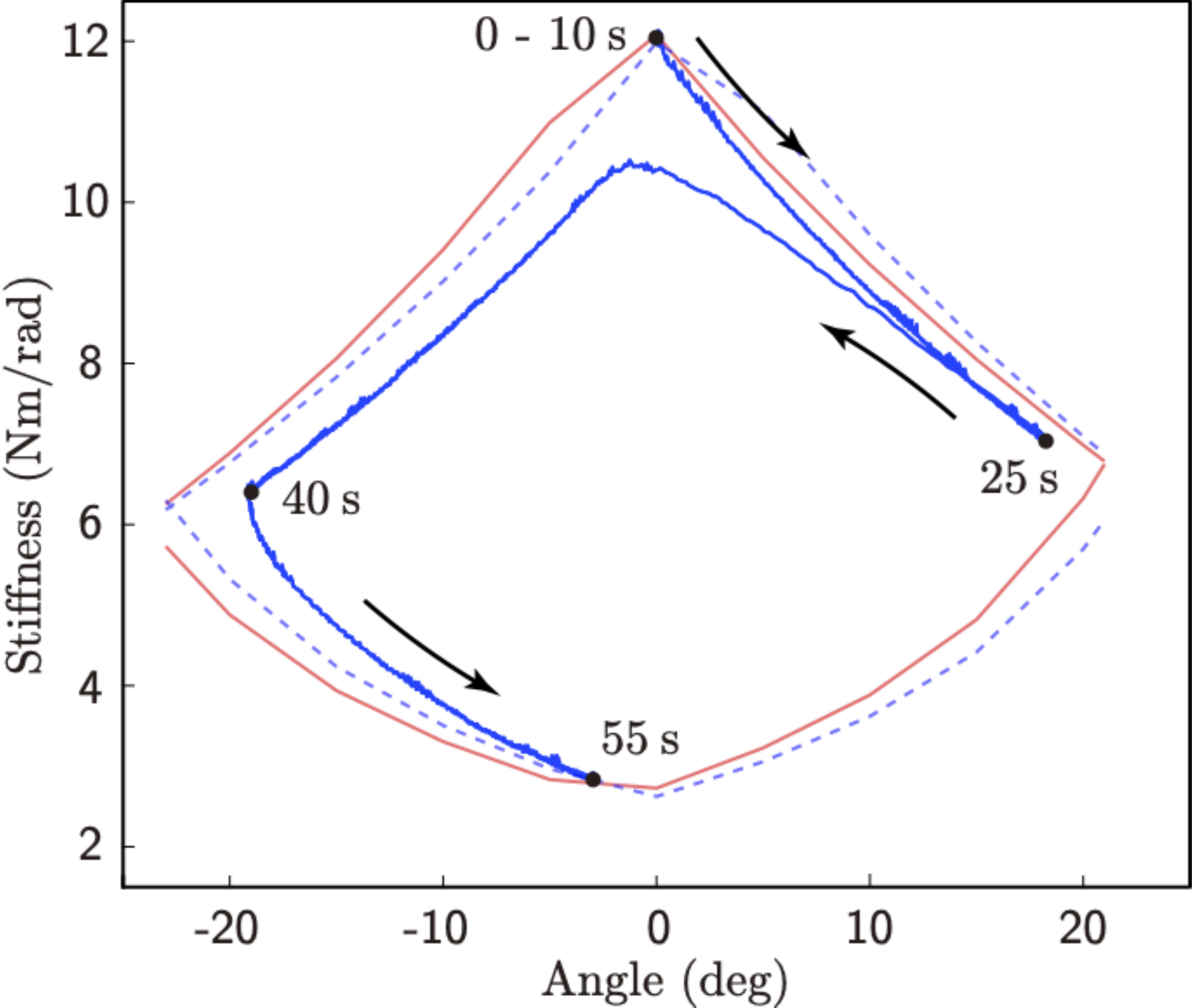}%
	\label{sfig:locus_exp}}
	\caption{Experimental results for angle and stiffness response.}
	\label{fig:valid}
\end{figure}

The reference set serves as a guide to set the reference values so that the experimentally obtained steady-state responses of a joint angle and stiffness remain within the reference set.
The following voltage signals ($u_1$ and $u_2$) are input to the PDCVs to cover the entire driving pressure range of the PAM (200 to 750 kPa) and obtain the time response of the joint angle and stiffness.
\begin{align*}
	u_1(t) = \begin{cases}
    		6 & 0\leq t\leq25 \\
    		4.7 &25< t\leq55
	\end{cases},
	u_2(t) = \begin{cases}
    		6 & 0\leq t\leq10 \\
    		4.5 &10<t\leq25\\
		6 & 25<t\leq40\\
    		4.5 &40<t\leq55
	\end{cases}.
\end{align*}
The time responses of the inner pressures of the two PAMs are shown in Figure.~\ref{fig:valid}\subref{sfig:pressure}, 
the time responses of the joint angle and stiffness are shown in Figure.~\ref{fig:valid}\subref{sfig:ak}, 
and the joint angle and stiffness trajectories on the admissible set are shown in Figure.~\ref{fig:valid}\subref{sfig:locus_exp}.
Focusing on the steady-state values of the joint angle and stiffness at 0 to 10, 25, 40, and 55 s, marked by black circles in Figure.~\ref{fig:valid}\subref{sfig:locus_exp}, all of the steady-state values of the joint angle and stiffness are clearly within the reference set.
Thus, this set can be used as an indicator for setting the reference joint angle and stiffness.

\subsection{Sensor-less Angle/Stiffness Control}
\begin{figure}[t]
		\centering
		\subfigure[Time responses of actual inner pressure of PAM1 (upper) and PAM2 (lower).]
		{\includegraphics[width=0.55\hsize]{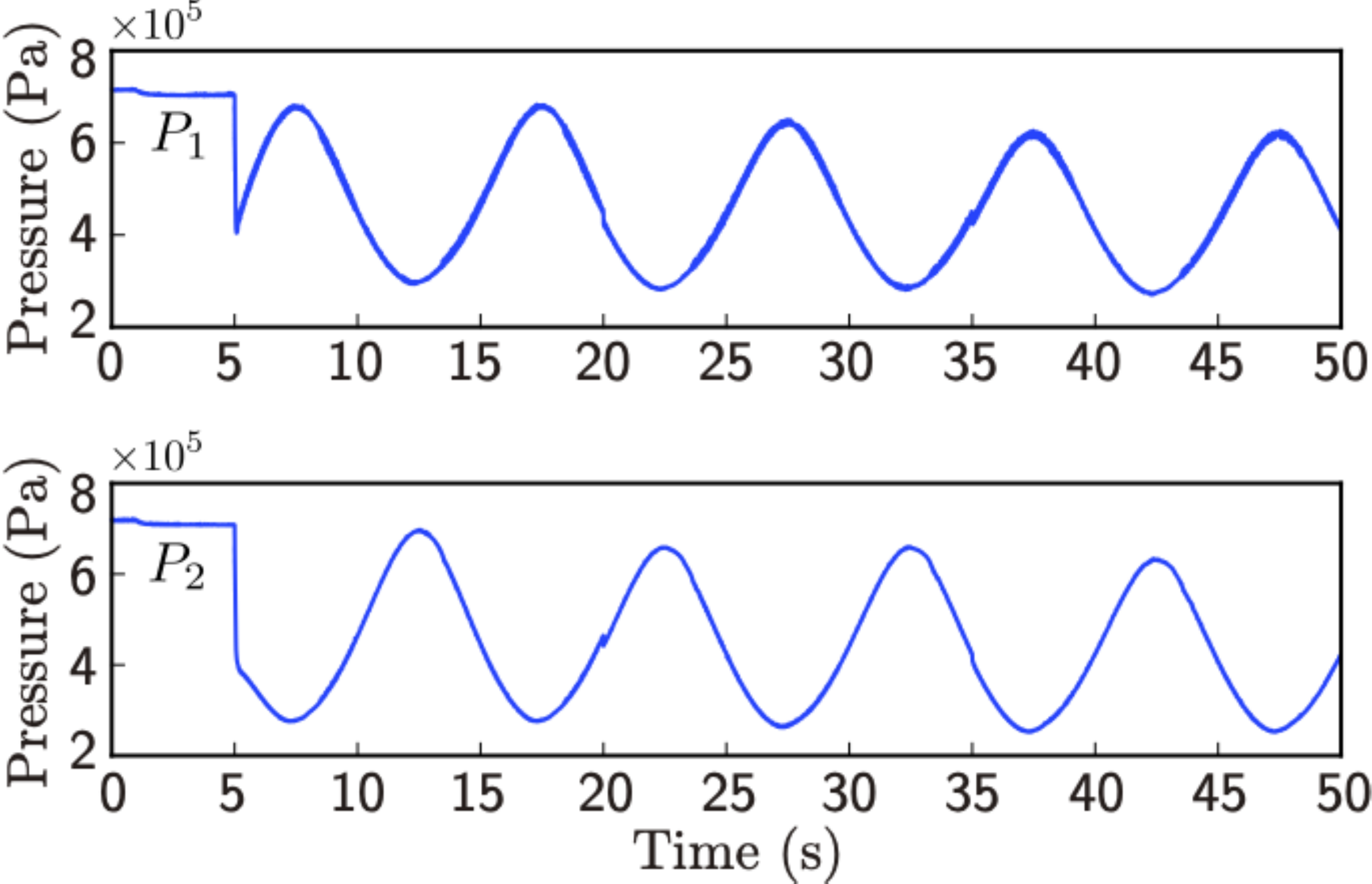}
		\label{subfig:pm15_P}}\\
		\subfigure[Time responses of joint angle (upper) and stiffness (lower).]
		{\includegraphics[width=0.55\hsize]{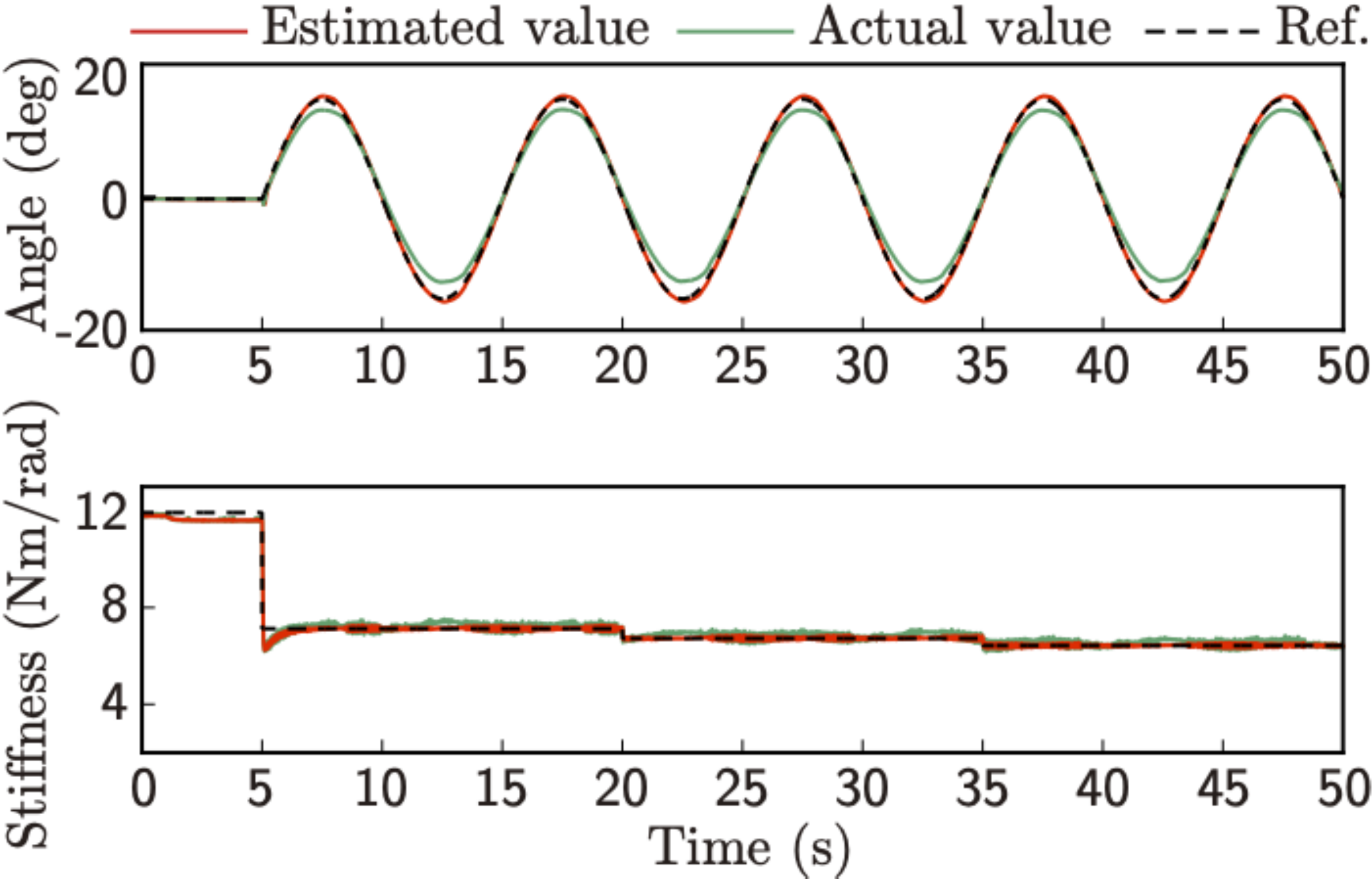}%
		\label{subfig:pm15}}\\
		\subfigure[Trajectories on the reference admissible set.]
		{\includegraphics[width=0.5\hsize]{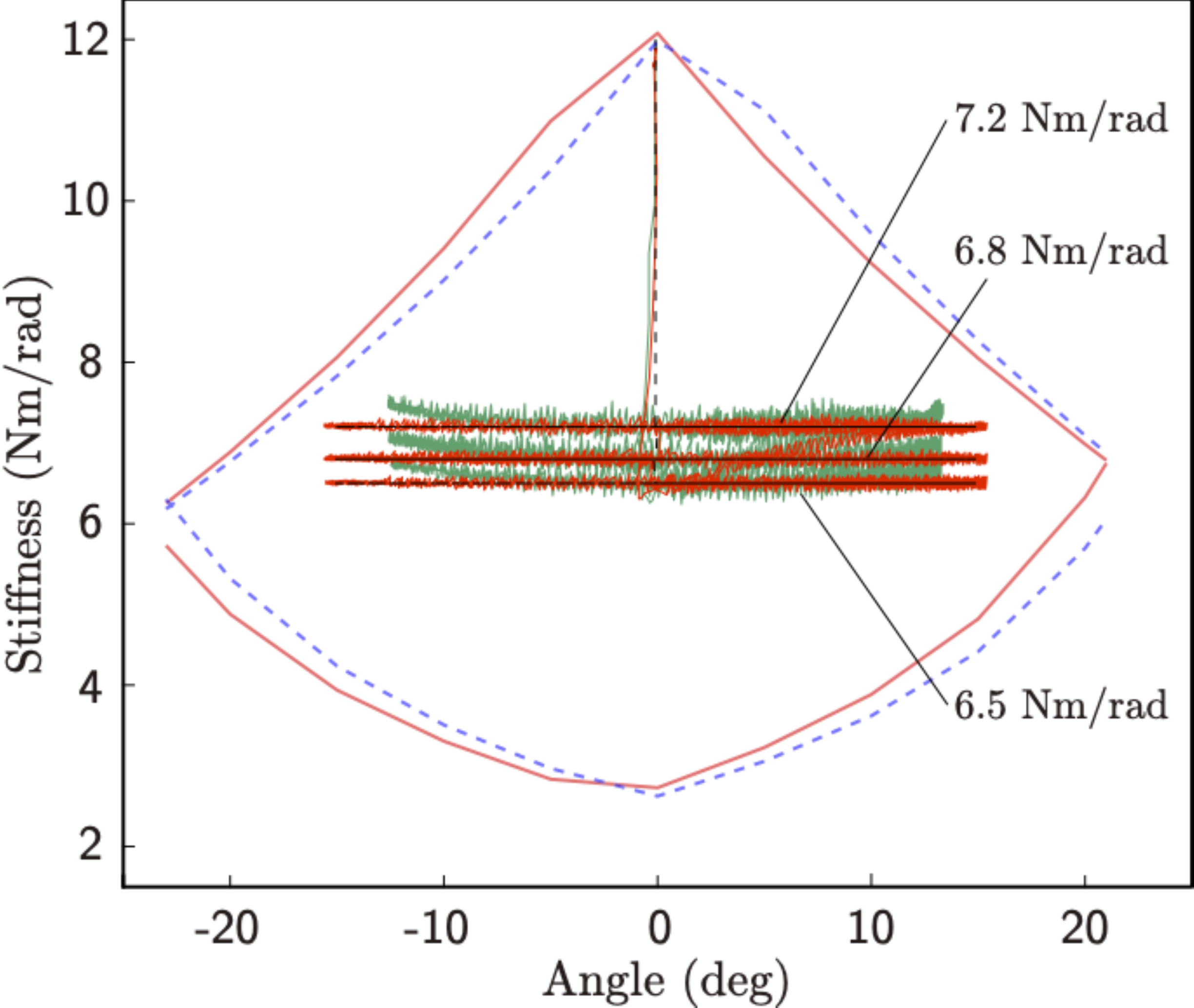}%
		\label{subfig:pm15_region}}
		\caption{Angle/stiffness control results (amplitude of reference angle: 15deg).}
		\label{fig:AK15deg}
	\end{figure}

\begin{figure}[t]
	\centering
	\subfigure[Time responses of actual inner pressure of PAM1 (upper) and PAM2 (lower).]
	{\includegraphics[width=0.55\hsize]{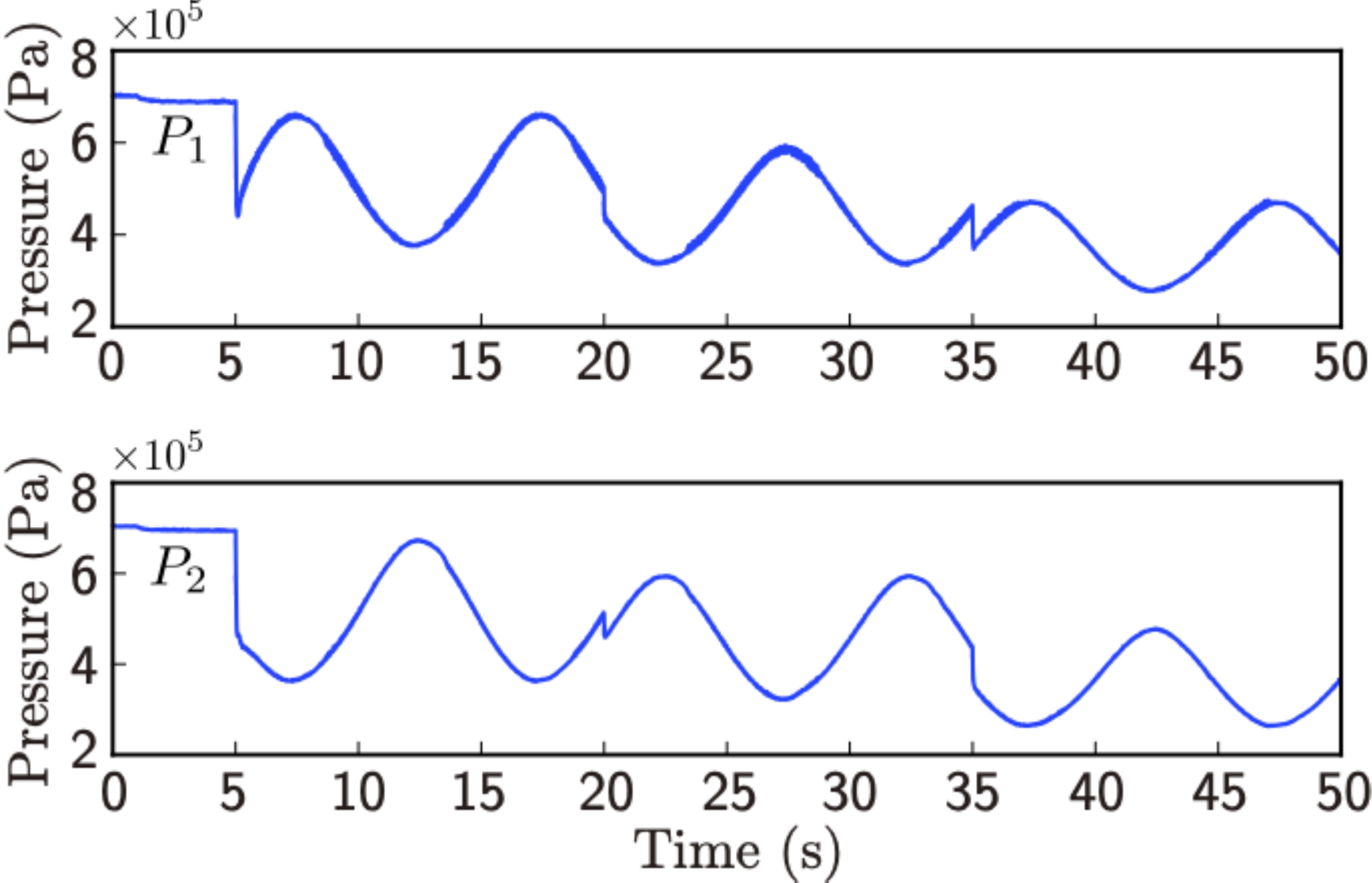}
	\label{subfig:pm10_P}}\\
	\subfigure[Time responses of joint angle (upper) and stiffness (lower).]
	{\includegraphics[width=0.55\hsize]{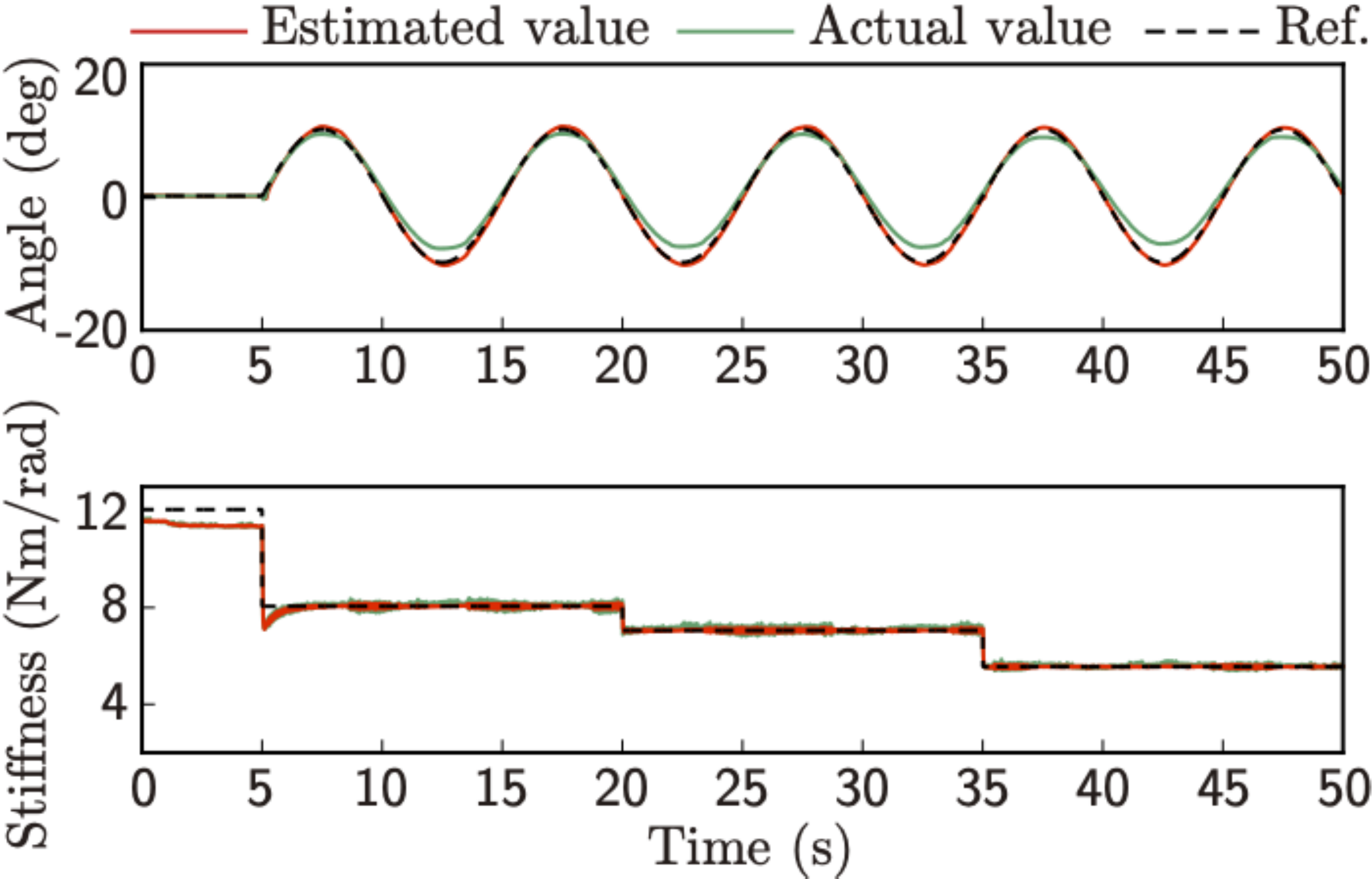}%
	\label{subfig:pm10}}\\
	\subfigure[Trajectories on the reference admissible set.]
	{\includegraphics[width=0.5\hsize]{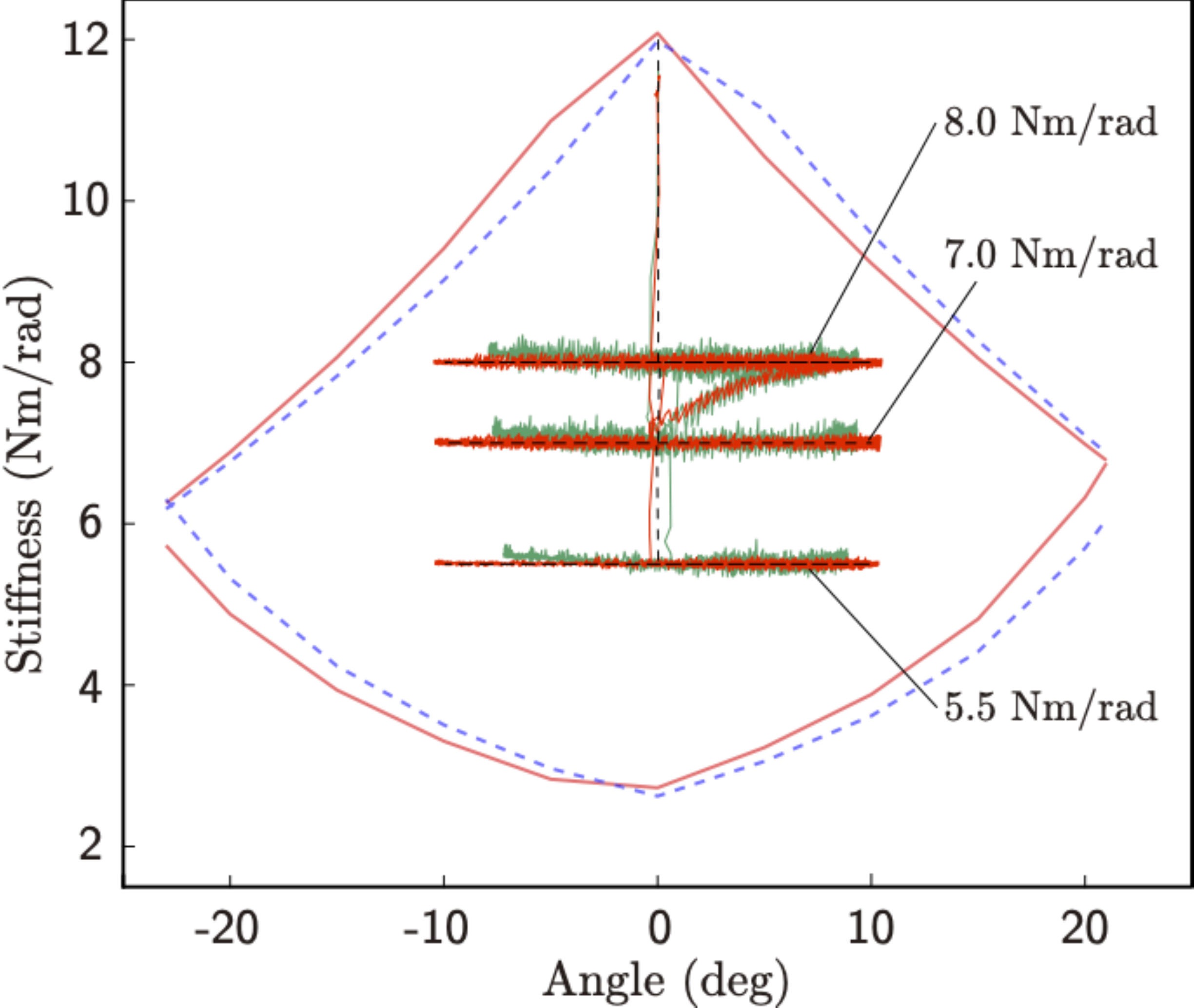}%
	\label{subfig:pm10_region}}
	\caption{Angle/stiffness control results (amplitude of reference angle: 10deg).}
	\label{fig:AK10deg}
\end{figure}

\begin{figure}[t]
		\centering
		\subfigure[Time responses of actual inner pressure of PAM1 (upper) and PAM2 (lower).]
		{\includegraphics[width=0.55\hsize]{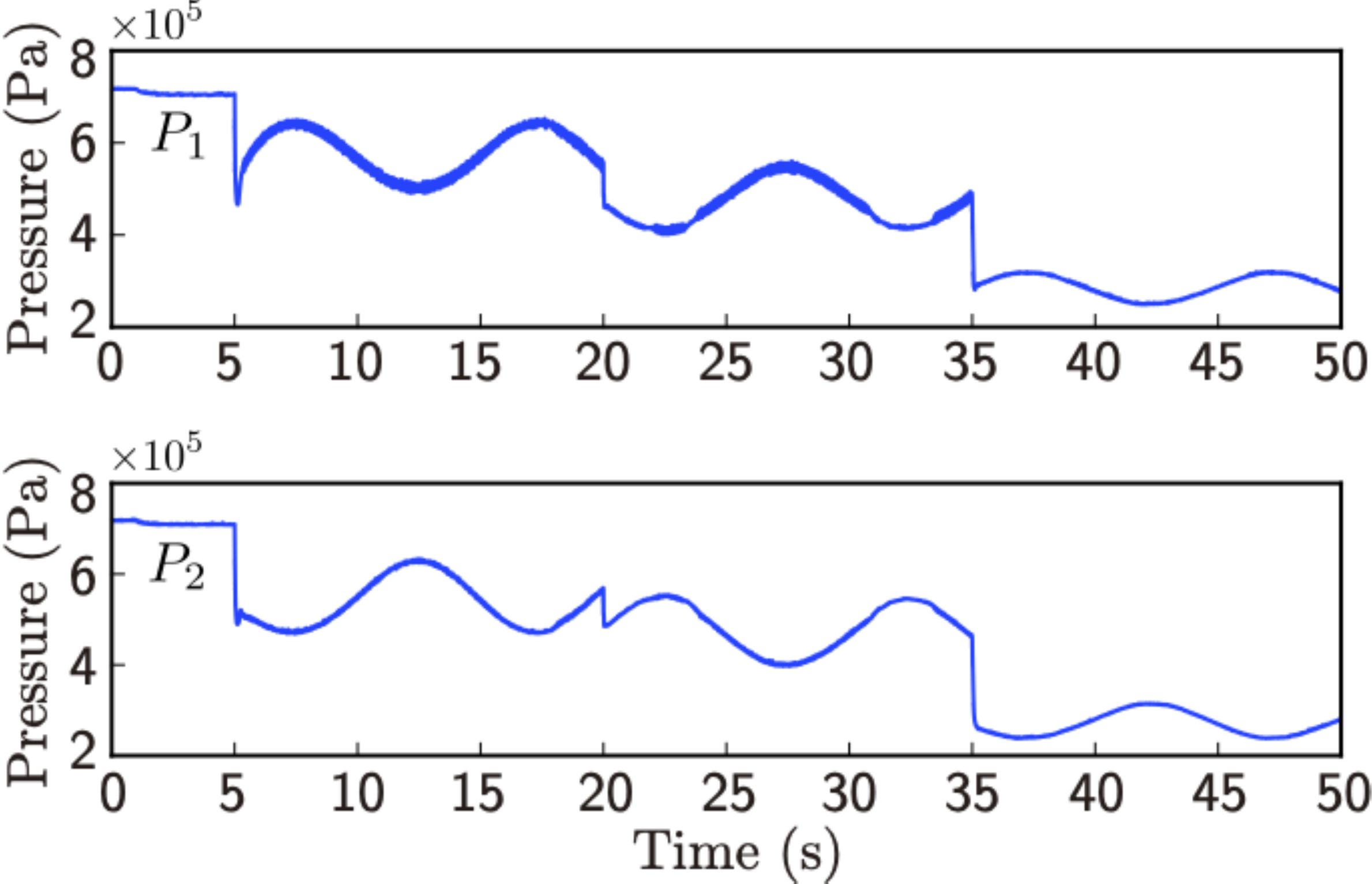}
		\label{subfig:pm5_P}}\\
		\subfigure[Time responses of joint angle (upper) and stiffness (lower).]
		{\includegraphics[width=0.55\hsize]{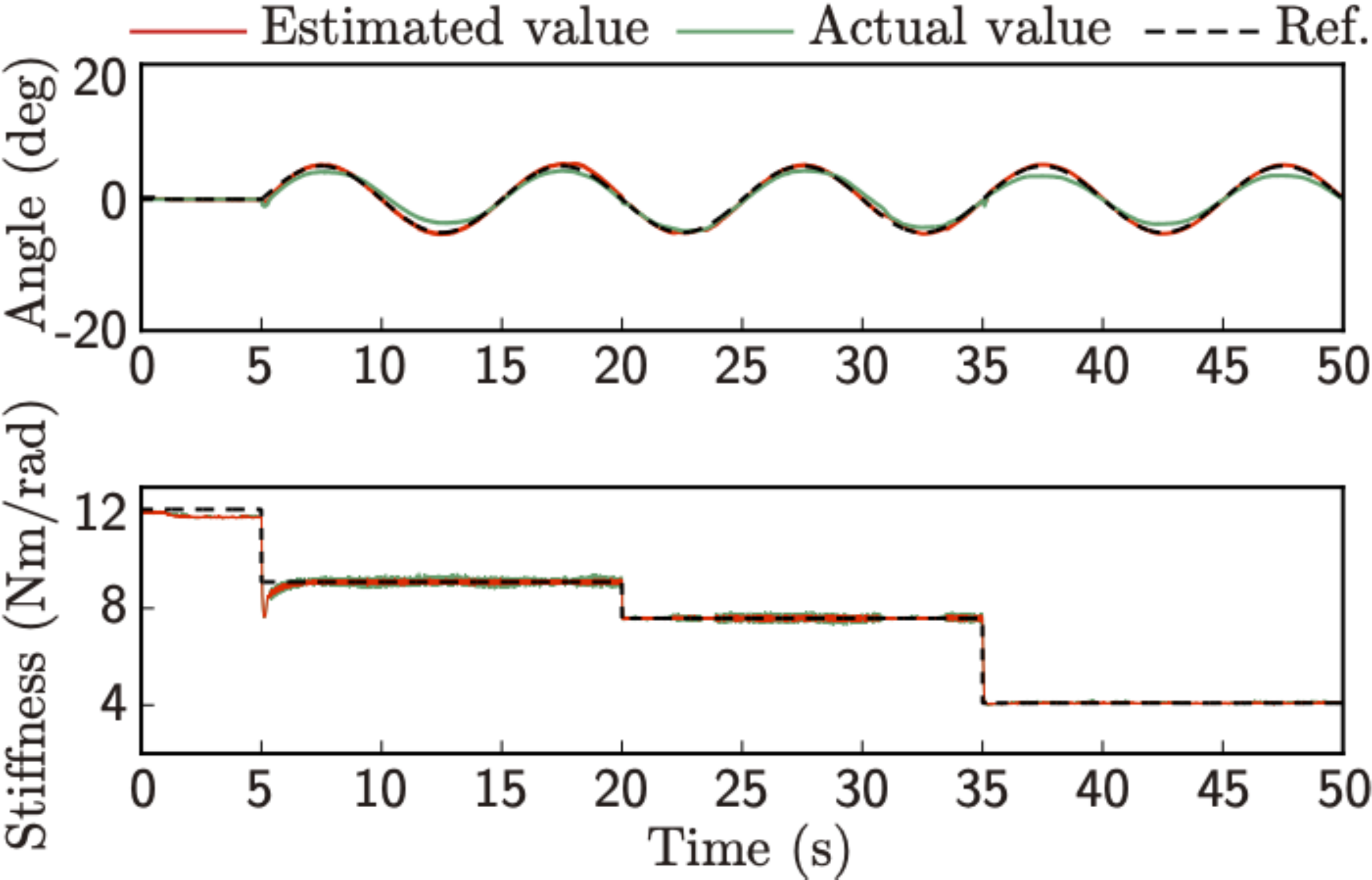}%
		\label{subfig:pm5}}\\
		\subfigure[Trajectories on the reference admissible set.]
		{\includegraphics[width=0.5\hsize]{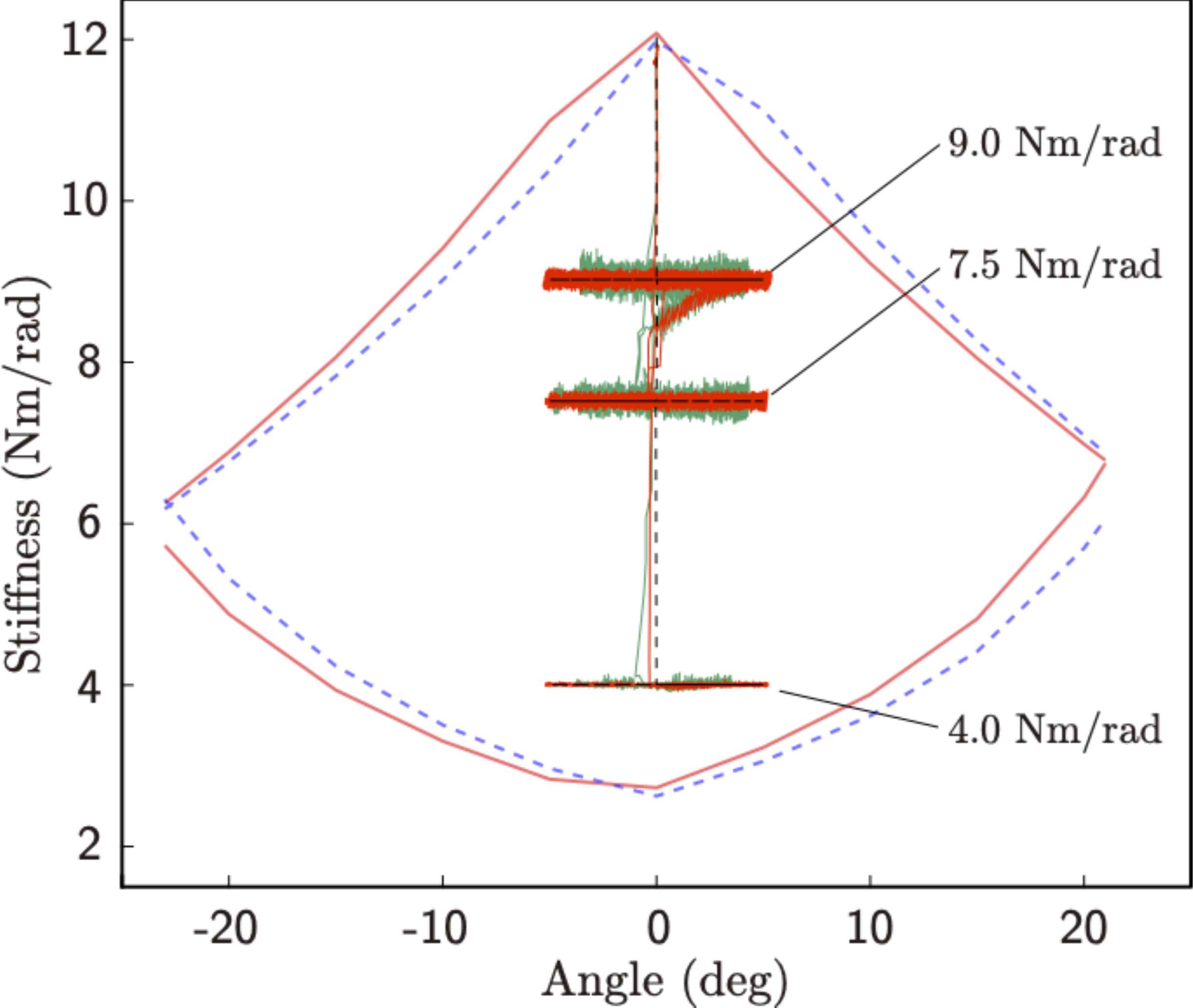}%
		\label{subfig:pm5_region}}
		\caption{Angle/stiffness control results (amplitude of reference angle: 5deg).}
		\label{fig:AK5deg}
	\end{figure}

Three experimental results for the proposed sensor-less angle/stiffness control are shown in Figures.\ \ref{fig:AK15deg} to \ref{fig:AK5deg}.
In each of the figures, 
(a) shows the time response of the pressure, 
(b) shows the time response of the joint angle and stiffness, and 
(c) shows the trajectory of the joint angle and stiffness on the computed admissible reference set.
In (b) and (c), the black dashed line is the reference, the solid red line represents the estimation based on the pressure sensor measurements, and the solid green line represents the actual value, computed by \eqref{eq:K} based on measurements using the pressure sensors and the encoder.

A reference of a joint angle was defined in each of the experiments in Figures.\ \ref{fig:AK15deg}, \ref{fig:AK10deg}, and \ref{fig:AK5deg} by a sinusoidal function with a period of 10 s and amplitudes set to 15, 10, and 5 deg, respectively.
To serve as a reference stiffness, three step-like signals within the computed set in Figure.~\ref{fig:region}\subref{sfig:region} were chosen with ranges of 7.2 to 6.5, 8 to 5.5, and 9 to 4 Nm/rad, respectively.
The proposed control method can be observed to accurately track the angle and stiffness to the references in all three figures.
Furthermore, it can be clearly observed in (c) of each figure that the angle/stiffness trajectories are within the reference admissible set.
These results confirm that the proposed method is capable of controlling the joint angle and stiffness independently without an encoder.

\section{Conclusion}\label{sec:conc}
This study proposed a sensor-less angle/stiffness control method for an antagonistic PAM actuator system and provided a procedure to obtain a set of admissible references, defined as pairs of stiffnesses and joint angles.
In order to realize the sensor-less control using only pressure measurements, this study applied a UKF within a detailed model to estimate the joint angle and contraction forces.
It was then demonstrated that the reference admissible set obtained using the model helps to choose an allowable reference. 
Three experiments were conducted using the characterized reference admissible set, and it was confirmed that the proposed method can control the stiffness and angle simultaneously and independently based only on the measured PAM pressures.
These experimental results indicate that the antagonistic PAM actuator is applicable to various devices that must be lightweight, low-cost, and interact safely with humans, such as nursing care robots, rehabilitation orthoses, and power-assist orthoses.

In future work, a control method for PAM-actuated devices will be developed considering the presence of disturbances such as reaction torque from the human arm.
This could realize safer operation of human-assisting robots by combining stiffness control with a disturbance observer.

\bibliographystyle{tADR}
\bibliography{reference}

\label{lastpage}

\end{document}